# Spin-based magnetic detection of optically trapped single cell in microfluidic channel

Jun Yin(殷俊)[1,2,3], Sanyou Chen(陈三友)[1,2,3], Yihao Yan(燕一皓) [1,4], Mengqi Wang(王孟祺) [1,2], Ya Wang(王亚) [1,2,4], Yiheng Lin(林毅恒) [1,2,4], Qi Zhang(张琪)[3,5†], and Fazhan Shi(石发展)[1,2,3,4†]

[1]School of Physical Sciences, University of Science and Technology of China, Hefei 230026, China
[2]Anhui Province Key Laboratory of Scientific Instrument Development and Application, University of Science and Technology of China, Hefei 230026, China
[3]School of Biomedical Engineering and Suzhou Institute for Advanced Research, University of Science and Technology of China, Suzhou 215123, China
[4]Hefei National Laboratory, University of Science and Technology of China, Hefei 230088, China
[5]Institute of Quantum Sensing, School of Physics, Institute of Fundamental and Transdisciplinary Research, Zhejiang Key Laboratory of R&D and Application of Cuttingedge Scientifc Instruments, State Key Laboratory of Ocean Sensing, Zhejiang University, Hangzhou 310027, China

Combining optical tweezers with fluorescence microscopy is a powerful tool for single-cell analysis, playing a pivotal role in disease diagnosis, cell sorting, and the investigation of cellular dynamics. However, fluorescence detection faces challenges such as blinking, photobleaching and autofluorescence in biotissues. To address these limitations, we developed a magnetic detection strategy by integrating quantum magnetometry using nitrogen-vacancy centers into optical tweezers, demonstrating precise trapping and manipulation of individual cells in microfluidic environment. We detected a magnetic signal of 89 μT from a single cell labeled with magnetic nanoparticles, compared to a noise floor of 3.9 μT observed in unlabeled cells. This platform provides a promising approach for high-precision single-cell analysis and holds significant potential for probing cellular activities within biological microenvironments.

## 1. Introduction

Fluorescent staining techniques enable selective labelling of specific components on cellular membrane or within intracellular compartments. Optical tweezers achieve non-contact, non-damage trapping of individual cells via laser-induced force. Combining these techniques allows real-time manipulation coupled with microscopic dynamic fluorescent tracking. This system has achieved high accuracy cell sorting[1-3] within microliter-scale samples, which is a critical step for early cancer diagnosis.[4] Spatial regulation of cell-substrate interactions, combined with simultaneous fluorescence measurement, advances drug testing[5] and addressing cancer cell chemotaxis in microenvironments.[6] However, fluorescence-based detection often suffers from background fluorescence, signal instability, light scattering and fluorescence blinking, which degrade the resolution and the sensitivity. Magnetic labeling-assisted single-cell analysis represents a promising alternative strategy because of near background-free property and signal stability. Magnetic detection techniques such as giant magnetoresistance (GMR), superconducting quantum interference (SQUIDs), and Hall sensors are capable of detecting magnetically labeled biological samples,[7] but face challenges in achieving nanoscale resolution.

---

† Corresponding author. E-mail: zhq2011@ustc.edu.cn
† Corresponding author. E-mail: fzshi@ustc.edu.cn

The nitrogen-vacancy (NV) center in diamond, serving as a quantum sensor, exhibits stable fluorescence, exceptional magnetic sensitivity, high spatial resolution, biocompatibility, and operability under ambient conditions. These properties make it an effective tool for nanoscale magnetic field measurements,[8, 9] demonstrating unique advantages in biomedical applications. NV center has been extensively applied to tumor tissue imaging,[10] biomolecular interaction detection,[11] biomagnetism sensing[12] and DNA assay.[13] Magnetic detection of single cells labeled with magnetic nanoparticles (MNPs) using NV centers has been realized.[14] However, such a platform lacks capabilities for real-time cell manipulation and sorting, which limits the application in single-cell studies.

Over the past decade, optical tweezers have demonstrated successful trapping of both ensembles and individual diamond particles in solution[15-23] and vacuum[24-26]. Through precise positional and angular[16, 21, 25] control of nanodiamonds, the optically detected magnetic resonance (ODMR) spectra and coherence times of NV centers have been measured. Optically trapped nanodiamonds have enabled investigations into torsional oscillations[25] and cooperatively enhanced dipole forces[20]. In addition, these levitated systems have demonstrated potential as sensors for mapping plasmonic antenna field densities[17] and detecting paramagnetic ions[19] in aqueous environments. However, the application of quantum diamond microscopy to studying optically trapped biological samples has remained largely unexplored. The development of instruments that combine optical tweezers with diamond quantum magnetometers lays the foundation for high-sensitivity single-cell analysis and sorting based on NV centers.

Here, we built a novel platform for single-cell manipulation and magnetic detection that integrates optical tweezers with NV centers, thereby introducing spin-based magnetic sensing to optically trapped single-cell studies. We employed single shallow NV centers in diamond as sensors, achieving a magnetic sensitivity of $3\ \mu T/\sqrt{Hz}$.

Through optical tweezers, individual MNP-labeled cells were stably trapped in a microfluidic chamber and precisely positioned over the diamond surface. By controlling the distance between the magnetic cell and the NV center, we observed a 2.48 ± 0.26 MHz shift in the ODMR frequency, corresponding to a magnetic field of 89 μT. The noise floor was 0.11 MHz measured with an unlabeled cell.

## 2. Basic theory and experiment set up

The integrated single-cell analysis platform combines quantum diamond microscopy with optical tweezers technology. As illustrated in **Fig. 1(a)**, a tightly focused 1064 nm laser beam generates optical force to trap and manipulate individual cells within a microfluidic channel. The sample chamber is mounted on a coplanar waveguide (CPW) integrated with a diamond chip. By precisely positioning cells near shallow NV centers in the diamond via optical tweezers, we could detect magnetic signals from labeled cells through ODMR.

The NV center in diamond consists of a substitutional nitrogen atom and a nearby vacancy, exhibiting $C_{3v}$ symmetry. This defect has two stable charge states,[27] i.e., the negatively charged $NV^-$ and the neutral $NV^0$, with the $NV^-$ typically employed for quantum sensing. The ground-state ($^3A_2$) electronic spin of $NV^-$ is a triplet state, and its energy-level structure is shown in **Fig. 1(b)**. In the absence of an external magnetic field, the $|m_s = \pm 1\rangle$ remain degenerate, whereas an applied magnetic field induces Zeeman splitting. The magnetic-field-dependent energy shift is quantified by:[28]

$$\Delta f = m_s \gamma_{NV} B_z, \qquad (1)$$

where $\gamma_{NV} = 2.8\,\text{MHz/G}$ is the gyromagnetic ratio of the NV center and $B_z$ denotes the magnetic field component along the NV axis. A 532 nm laser is commonly used to initialize and excite the NV center. The transition from the excited state back to the ground state is spin-dependent: $|m_s = \pm 1\rangle$ preferentially relax via intersystem crossing to the singlet state, while $|m_s = 0\rangle$ has higher probability of directly retuning to the ground state with fluorescence emission, resulting in stronger fluorescence than $|m_s = \pm 1\rangle$. Spin-dependent relaxation enables optical initialization and readout of spin state. Microwave-driven spin transitions are probed by sweeping the microwave frequency and recording fluorescence intensity from NV centers, thereby resolving spin-resonance frequencies that form the basis of ODMR.

A mixed microliter-scale sample of MNP-labeled target cells (e.g., cancer cells) and unlabeled control cells is loaded into the microfluidic chip. An optical field with an intensity gradient exerts a force on microscopic objects, capturing them toward regions of higher intensity. We utilize this gradient force to trap cells and manipulate their positions by controlling the focal point of the 1064 nm laser. When positioned near the NV center, target cells labeled with MNPs generate magnetic fields at the NV center [**Fig. 1(c)**, right], inducing a shift in ODMR spectral peak position. In contrast, unlabeled control cells [**Fig. 1(c)**, left] show undetectable spectral shifts. The presence or absence of this resonant frequency shift enables the distinction between target and non-target cells. Following ODMR-based magnetic detection, optical tweezers enable the transport of the analyzed cells to designated regions, which enables the extraction of target cells from test samples for subsequent research.

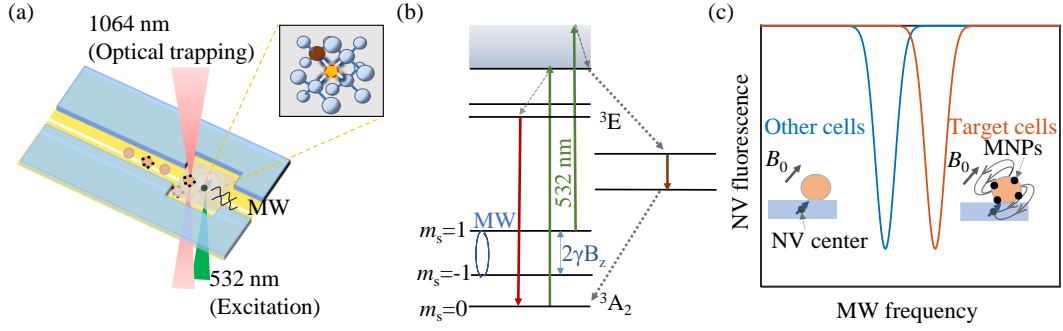

**Fig. 1.** Schematic diagram of single-cell detection. (a) A diamond chip containing shallow NV centers is positioned within the microfluidic channel to serve as a magnetic sensor. The lattice structure of NV center is shown in the inset. A 532 nm laser and microwave radiation are applied to the NV center for spin manipulation. A tightly focused 1064 nm laser beam provides gradient force to trap and manipulate individual cells. (b) Energy level structure of the NV$^-$. The energy levels exhibit Zeeman splitting under an external magnetic field. (c) The magnetic field produced by MNP-labeled cells causes a shift in the ODMR resonance peak corresponding to transition between $|m_s = 0\rangle$ and $|m_s = 1\rangle$, whereas no shift will be observed with blank cells.

The experimental setup is illustrated in **Fig. 2**. The core equipment is a home-built confocal microscope, consisting of optical tweezers, an ODMR module, and a microfluidic sample chamber. The 1064 nm laser and the 532 nm laser are modulated by separate acousto-optic modulators (AOM) (Isomet M1250-T220L-0.5 for 532 nm, Isomet M1099-T80L-3 for 1064 nm) to generate laser pulses with a rising time of ~200 ns. The 1064 nm laser beam (Changchun New Industries MGL-III-1064-1W), after fiber filtering, is overlapped with the 532 nm laser via a pair of mirrors. A fast steering mirror (FSM, Newport FSM-300) enables lateral scanning of the 1064 nm light focus. The 1064 nm laser focus is scanned along the optical axis by changing the focal power of a tunable lens (EL, Optotune EL-10-30-C). A telescope, formed by two confocal lenses with 50 cm focal length, conjugates the FSM and the back focal plane of the objective. Both the 532 nm beam (Changchun New Industries MGL-III-532-150 mW) and the 1064 nm beam are focused by an objective with 1.45 NA (Nikon Plan Apo λ, 100× oil). The fluorescence from the NV centers is collected by the objective and detected by a single-photon avalanche diode (APD, Excelitas SPCM-AQRH). Dichroic mirror 1 in the optical path couples the 532 nm laser with the 1064 nm laser, and dichroic mirror 2 separates the 1064 nm light from the NV fluorescence. An 8:92 (reflection: transmission) pellicle beam splitter (Thorlabs BP108) directs widefield illumination to a charge-coupled device (CCD, Andor DU-897U-CS0-EXF) camera for real-time sample monitoring.

The diamond chip (2 mm × 2 mm × 0.05 mm), mounted on CPW, is implanted with $N_2^+$ at 8 keV beam energy and with a dose of 1×10$^9$ nitrogen atoms/cm$^2$. The expected NV centers depth ranges between 5 and 10 nm. A microfluidic chip is adhered to CPW, which radiates microwaves to manipulate the spin state. To achieve sufficiently high microwave intensity and a spatially uniform microwave field distribution, we employed an Ω-shaped CPW with a central optical aperture of 1 mm in diameter. The entire sample stage is mounted on a 3D piezo nanopositioner (Physik Instrumente P-562.3CD) for nanometer-precision movement. The FSM enables lateral scanning of the optical trap across a 53 μm × 65 μm range, limited by the dimensions of the optical components. In parallel, the sample chamber, integrated with a hybrid positioning system, which comprises a piezo nanopositioner (1 nm resolution) and a translation stage (millimeter-scale travel range), enabling coarse-to-fine sample navigation across multiple spatial scales.

The control circuits of the instrument are shown in **Fig. 2(b)**. A pulse generator (PulseBlasterESR-PRO 500MHz, PB) controls the pulse sequences via TTL signals, to complete the timing operations of laser and microwave. In these circuits, a PIN switch is used to switch the microwave (MW) to general MW pulses. The PB also triggers two AOMs to modulate the 532 nm and the 1064 nm laser pulses, respectively. The deflection angle of the FSM is determined by the applied piezoelectric voltage. The timing operations, which comprise the movement of the optical trap center and ODMR detection, are programmatically generated to synchronize all subsystems.

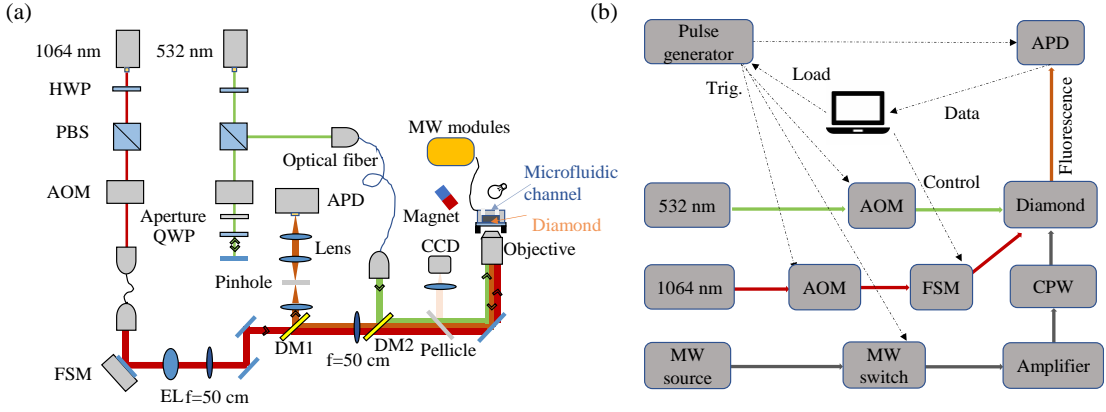

**Fig. 2.** Experiment setup. (a) Optical layout of the platform. HWP, half-wave plate. PBS, polarizing beamsplitter. AOM, acousto-optic modulator. FSM, fast steering mirror. EL, electrically focus tunable lens. QWP, quarter-wave plate. DM, dichroic mirror. APD, single-photon avalanche diode. CCD, charge-coupled device camera. (b) Hardware control of experimental equipment.

To enhance the magnetic sensitivity of NV centers, we employed a permanent magnet to apply a static magnetic field of approximately 400 G along the NV axis which polarizes the nitrogen nuclear spin. This external field also aligns the magnetic moments of MNPs. Pulsed optically detected magnetic resonance (pulsed-ODMR) serves as an effective method for high-sensitivity DC magnetic field measurements. As illustrated in **Fig. 3**, the $NV^-$ spin state is first optically initialized to $|m_s=0\rangle$ state. Subsequently, a resonant microwave π-pulse is applied to flip the spin population. Fluorescent photons are counted during reference (Ref. in **Fig. 3**) and signal (Sig. in **Fig. 3**) acquisition windows. The fluorescence (*FL*) contrast is defined as:

$$\text{contrast} = \frac{FL_{\text{Sig.}} - FL_{\text{Ref.}}}{FL_{\text{Ref.}}}. \qquad (2)$$

Normalizing against the reference photon counts compensates for errors induced by laser power fluctuations or platform drift. The 1064 nm laser can ionize $NV^-$ from the excited state to $NV^0$,[29, 30] which has been applied for background-free imaging.[31] This charge state transitions degrade spin polarization of $NV^-$, thereby decreasing the FL contrast when the 1064 nm laser remains "on" during ODMR detection [**Fig. 3(b)** and black data in **Fig. 3(c)**]. To overcome this issue, we employed an alternating 1064 nm (Alt-1064 nm) laser [**Fig. 3(a)**] instead of simultaneous (Sim-1064 nm) laser for optical trapping. The 1064 nm laser operates with a 2:1 duty cycle (on:off), and ODMR detection is carried out during the off intervals of 1064 nm, with the experimental signal accumulated over multiple repetitions. This intermittent trapping protocol maintained stable cell confinement within the optical tweezer. Replacing simultaneous light with an alternating 1064 nm laser increased the ODMR contrast from 17% to 33% [red data in **Fig. 3(c)**].

The DC magnetic sensitivity of pulsed-ODMR is given by:[32]

$$\eta = \frac{Fh\Delta v}{g\mu_B C\sqrt{N}} \qquad (3)$$

where $F$ is the ODMR line shape fitting parameter ($F=0.70$ for Gaussian and $F=0.77$ for Lorentzian line shapes), $h$ represents the Planck constant, $g = 2.0028$ is the g-factor of the NV center, $\mu_B$ is the Bohr magneton and $\Delta v$ is the full width at half maximum (FWHM) of the spectral peak, which depends on the microwave field strength and the spin coherence time of the NV center. $C$ is the resonance contrast, and $N$ is the photon count rate, which is proportional to NV luminescence intensity and the collection efficiency. Excessive MW power induces power broadening of the resonance linewidth, while insufficient power fails to fully drive spin transitions. Optimizing the MW power can strike a balance between a sufficiently strong Rabi drive (high contrast) and a narrow linewidth (high resolution), ensuring optimal sensitivity. Environmental factors, such as dielectric constant variations and electromagnetic noise at solution-immersed diamond surfaces, modulate the coherence time and π-pulse duration. Therefore, we optimized the DC magnetometry sensitivity of the NV centers in a cell solution sample. By optimizing the microwave power, we achieved a magnetometry sensitivity of $3\ \mu T/\sqrt{Hz}$ with $T_2^*=2$ μs, shown in **Fig. 3(d)**.

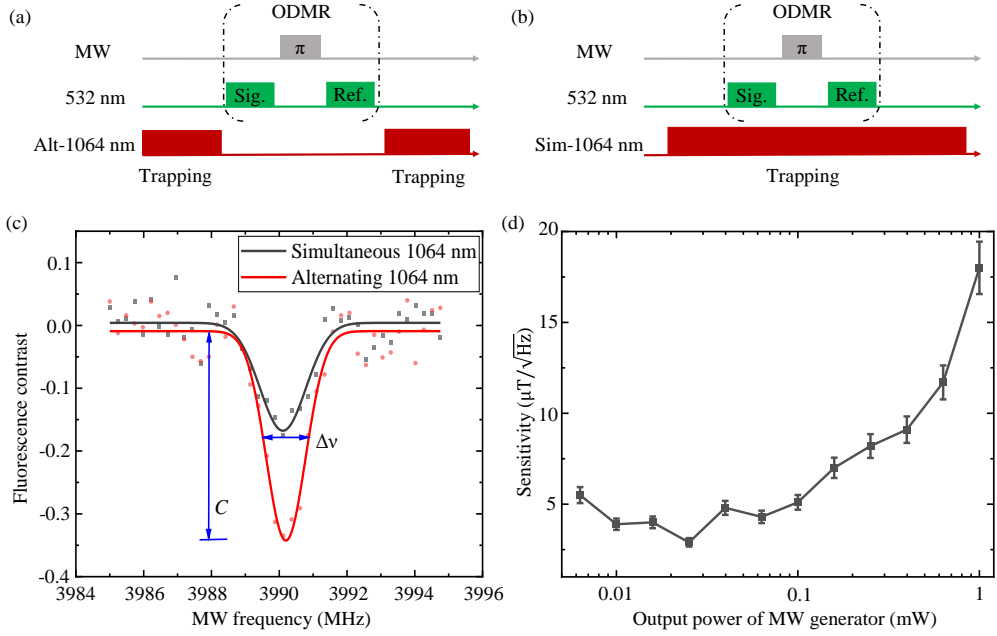

**Fig. 3.** Experimental sequence and ODMR spectrum for DC magnetic field measurement. (a) Pulsed-ODMR sequence with alternating trapping laser. (b) Pulsed-ODMR sequence with simultaneous trapping laser. (c) ODMR spectrum of a single NV center with an external magnetic field. The spectral peak corresponds to the transition $|m_s=0\rangle \to |m_s=+1\rangle$. (d) Optimization the magnetic detection sensitivity by sweeping the microwave power.

## 3. Result and discussion

In this study, J774.1 murine macrophage cells were labeled with internalized 20-nm $Fe_3O_4$ MNPs via endocytosis. During the cell culture phase, macrophages were maintained in culture medium supplemented with 10% fetal bovine serum (FBS). Following endocytosis, the cells were washed with phosphate-buffered saline (PBS) to remove non-internalized particles. Transmission electron microscopy (TEM) image (**Fig. 4**) revealed that MNPs were predominantly enriched in endosomes and lysosomes within the cytoplasmic compartment. While endocytosis-driven internalization was utilized here, we note that antigen-targeted MNP labeling via immunoreactions is a well-established alternative.[10, 14]

We simulated the magnetic signal patterns from a MNPs-labeled cell detected by the NV center [**Fig. 4(d)**]. The magnetic cell was approximated as a point dipole located at the center of the cell[14], with a magnetic moment $m = N \times m_{MNP}$ with $N = 4.5 \times 10^5$ MNPs and $m_{MNP} = 8.6 \times 10^{-16}$ emu for each MNP. The value of N was derived for a 15-μm-diameter cell, maintaining the same MNP density as shown in Fig. 4(c).

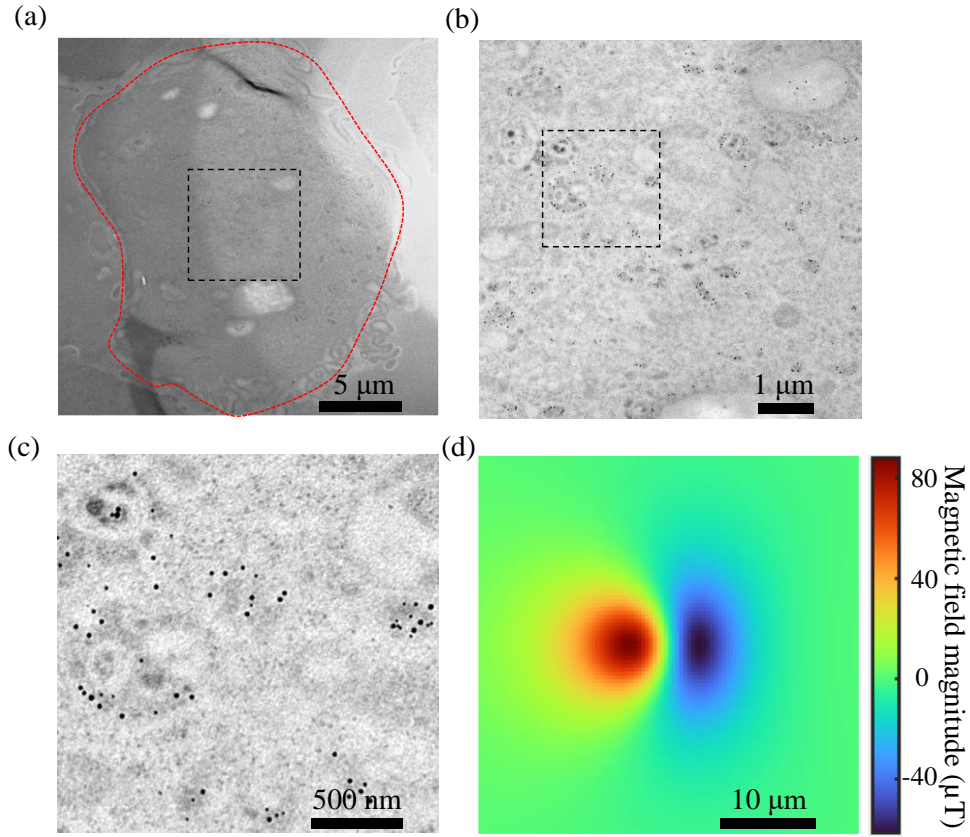

**Fig. 4.** (a)-(c) TEM image of murine macrophages labeled with MNPs. The red dashed line indicates the contour of sectioned cell. (b) and (c) are magnified views of the regions within the black dashed boxes in (a) and (b), respectively. (d) Simulated magnetic field magnitude at the NV center with MNP-labeled cell at different locations.

The cell suspension in PBS was loaded into the sample chamber through a microfluidic channel. To optimize the fluorescence stability and coherence properties of shallow NV centers, the surface of the diamond chip was functionalized with carboxyl groups. Under these conditions, cells in standard-concentration PBS rapidly adhered to the diamond surface,[33] and the axial gradient force provided by a 30 mW 1064 nm laser proved insufficient to displace the immobilized cells. To resolve this issue, PBS was diluted by a factor of 25, which effectively suppressed cell adhesion within 2 hours. Under these conditions, the cells retained their intact morphology.

Magnetic signal measurements were performed on optically trapped single cells. To minimize the heating effect of the laser, we reduced the power of the 1064 nm laser to 12 mW. Macrophages labeled with MNPs were cyclically translocated between two spatial positions (proximal and distal) separated by ~13 μm, and the ODMR spectra under the two configurations are shown in **Fig. 5(a)** and **5(b),** and the corresponding CCD images are displayed in **Fig. 5(c)**. In the proximal configuration, the NV sensor was 600 nm from the optical trap center. The experiment was repeated five times, the square wave in **Fig. 5(d)** represents the y-axis driving voltage of the FSM, with an amplitude of 1 V. After each translocation, a 5-second stabilization period ensured positional equilibrium of the trapped cell. ODMR measurements were then performed to resolve the magnetic field generated by the cell. All ODMR resonance frequencies obtained from repeated experiments are presented in **Fig. 5(d)**.

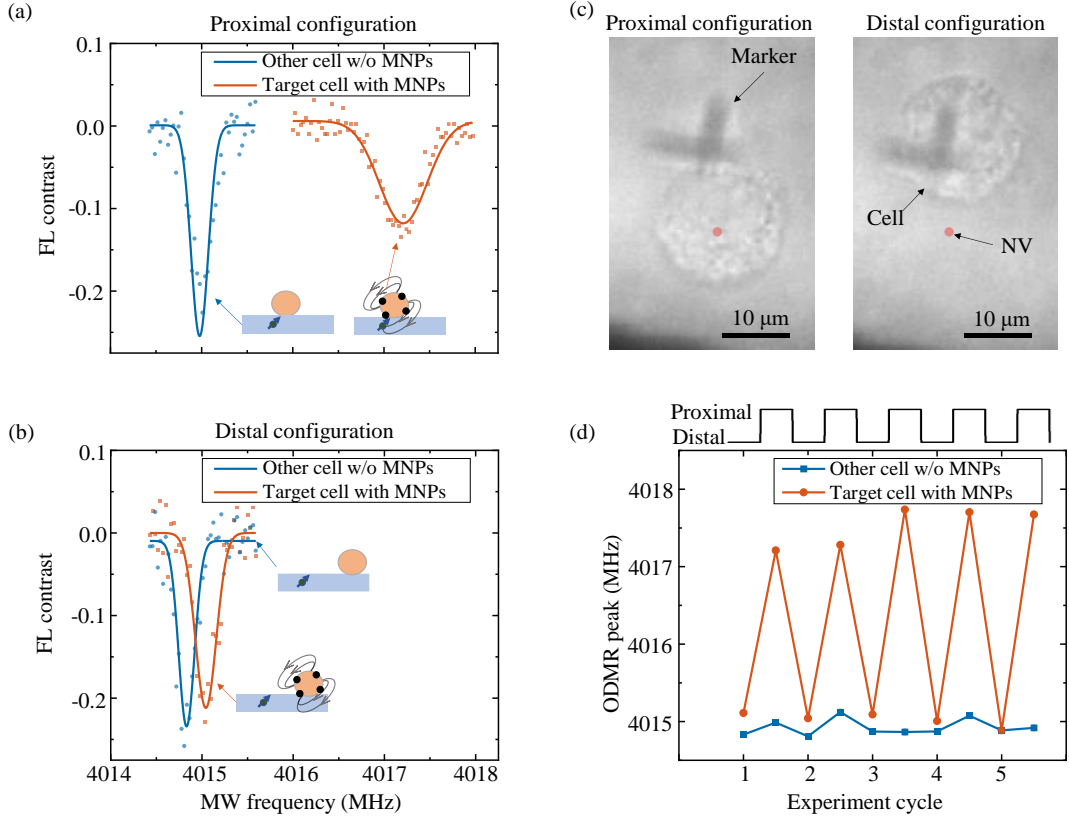

**Fig. 5.** Magnetic detection of optically trapped single cell with ODMR. (a) ODMR spectra of an MNP-labeled cell and an unlabeled cell in the proximal configuration. (b) The same with (a) but in the distal configuration. (c) Spatial manipulation of single cell on the diamond. The cross pattern is a marker on the diamond surface for positioning. (d) Top: the square wave represents the applied y-axis voltage of FSM. Bottom: resonance frequency of NV spin response to positional changes of the MNP- labeled cell and control cell.

Experimental results with MNP-labeled single cells demonstrated a resonance peak shift of 2.48 MHz between proximal and distal configurations ($f_{prox}$=4017.52, standard deviation: 0.26 MHz and $f_{dist}$=4015.04, standard deviation: 0.09 MHz), corresponding to an axial magnetic field variation of 89 µT. In contrast, control experiments using unlabeled cells exhibited no significant spectral shift ($f_{prox}$ =4014.99, standard deviation: 0.11 MHz and $f_{dist}$=4014.85, standard deviation: 0.03 MHz), confirming the observed magnetic signals originating exclusively from the MNP-labeled single cell.

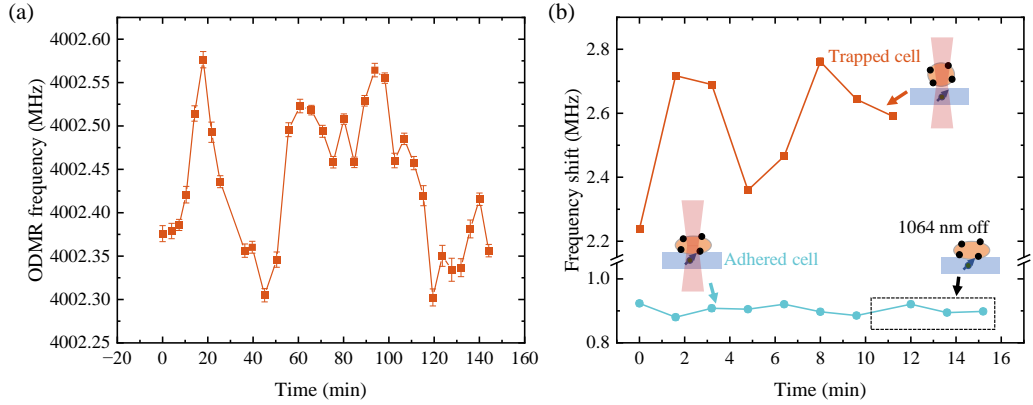

Fig. 6. (a) Time trace of the ODMR frequency without cells in microchannel shows the magnetic field instability of the system. (b) Time-resolved magnetic signals from an optically trapped cell (orange) and an adhered cell (cyan). For the data of the adhered cell, the last three points were recorded with the 1064 nm laser turned off. Error bars are obscured by data markers in the figure owing to fitting uncertainties below 0.01 MHz.

Fig. 6(a) characterizes external magnetic field fluctuations, showing a drift of 0.3 MHz over 150 minutes with an instability of 0–0.03 MHz/min. This instability matches well with the spectral variations of 0.11 MHz in the control experiments recorded within 50 minutes. For adhered cells [**Fig. 6(b),** cyan data], the signal instability (0.03 MHz/min) closely matched this baseline drift. In contrast, optically trapped cells [**Fig. 6(b)**, orange data] exhibited significantly higher drift rates (up to 0.3 MHz/min). These results are consistent with the greater spectral signal fluctuations observed in proximal magnetic cells compared to distal cells in **Fig. 5**. The enhanced instability originates from cellular deformation during optical manipulation and rotational motion within the trap. These findings suggest that NV-based magnetic sensing could serve as a novel probe for real-time monitoring of cellular deformation dynamics in future mechanobiology studies.

To evaluate potential laser-induced heating effects, we analyzed ODMR spectra from an NV center directly beneath an adhered cell under two conditions: (i) laser cycling (12 mW, 20 μs on and 10 μs off) and (ii) laser off. No significant resonant frequency shift was detected after switching off the laser. Thus, the laser-induced heating effects should be smaller than the baseline noise level of adhered magnetic cell (0.043 MHz peak-to-peak).

## 4. Conclusion and perspectives

We report a spin-based platform combining single-cell optical manipulation with quantum magnetic detection. The system integrates optical tweezers with ODMR sensing, achieving an optimized magnetic sensitivity of 3 $\mu T/\sqrt{Hz}$ using a single NV center. We experimentally demonstrated the capability of the platform to trap and manipulate individual cells within a microfluidic chamber. By comparing the signals measured in MNP-labeled cells and unlabeled control cells, spin-based magnetic sensing of single cell is displayed in this work. Our magnetic labeling strategy offers distinct advantages, including reduced background, enhanced signal stability, and immunity to optical field distortions. By using APD array detector or CCD camera to detect fluorescence from NV center array, this approach enables high-throughput flow cytometry and cell sorting. Leveraging the precise manipulation capabilities of optical tweezers, this platform holds potential for further investigation of cell-cell and cell-microenvironment interactions. Additionally, optical tweezers enable *in vivo* cellular manipulation[34] and intracellular organelle[35] control. Integrating ODMR sensing into optical tweezers paves the way for *in situ* detection of biological processes within living organisms.

**Acknowledgment**

This work was supported by the National Key R & D Program of China (Grant Nos. 2019YFA0709300, 2021YFB3202800), the National Natural Science Foundation of China (Grant Nos. T2125011, 12174377), the CAS (Grant Nos. YSBR-068), Innovation Program for Quantum Science and Technology (Grant Nos. 2021ZD0302200, 2021ZD0303204), New Cornerstone Science Foundation through the XPLORER PRIZE, Science and Technology Department of Zhejiang Province (2025C01041) and the Fundamental Research Funds for the Central Universities (226-2024-00142).